\documentclass[twocolumn,superscriptaddress,groupedaddress,10pt]{article}  

\usepackage{graphicx}  
\usepackage{dcolumn}   
\usepackage{bm}        
\usepackage{amssymb}   
\usepackage{amsmath}
\usepackage{authblk}
\usepackage[a4paper,margin=20mm]{geometry}
\usepackage[font=small]{caption}

\begin{document}

\title{ High-speed ultrasound imaging in dense suspensions reveals impact-activated solidification due to dynamic shear jamming}
\author[1]{Endao Han}
\author[1,2]{Ivo R. Peters}
\author[1]{Heinrich M. Jaeger} 
\affil[1]{James Franck Institute, The University of Chicago, Chicago, Illinois 60637, USA}
\affil[2]{Engineering and the Environment, University of Southampton, Highfield, Southampton SO17 1BJ, UK}

\date{\today}

\twocolumn[
  \begin{@twocolumnfalse}  
\maketitle

\textbf{
A remarkable property of dense suspensions is that they can transform from liquid-like at rest to solid-like under sudden impact.  Previous work showed that this impact-induced solidification involves rapidly moving jamming fronts; however, details of this process have remained unresolved.  Here we use high-speed ultrasound imaging to probe non-invasively how the interior of a dense suspension responds to impact.  Measuring the speed of sound we demonstrate that the solidification proceeds without a detectable increase in packing fraction, and imaging the evolving flow field we find that the shear intensity is maximized right at the jamming front. Taken together, this provides direct experimental evidence for jamming by shear, rather than densification, as driving the transformation to solid-like behavior. Based on these findings we propose a new model to explain the anisotropy in the propagation speed of the fronts and delineate the onset conditions for dynamic shear jamming in suspensions. \\
}
~\\
 \end{@twocolumnfalse}
]
\clearpage

Dense suspensions are complex fluids that can exhibit strong, discontinuous shear thickening (DST), where the viscosity jumps up orders of magnitude when a critical shear stress is exceeded \cite{Barnes,Seto,Cates_2,DST_review}. Under a wide range of dynamic conditions, dense suspensions can also undergo a transformation to solid-like behavior, for example during sudden impact at their free surface \cite{Scott,Stone,Shomeek}, ahead of quickly sinking objects \cite{Zhang,vonKann}, or during rapid extension \cite{Cates_3}. Detailed investigation of the dynamics during impact has shown how such solidification is associated with a propagating front that converts fluid-like, unjammed suspension into rigidly jammed material in its wake \cite{Scott, Ivo}. This dynamic jamming front moves through the suspension with a speed much greater than the impactor itself. 

To explain this solidification, a model was proposed \cite{Scott} that assumed the impact pushes the particles closer together until they jam. This densification scenario was based on the standard jamming phase diagram for frictionless hard particles, where entry into a jammed state requires an increase in particle packing fraction $\phi$ \cite{jamming,SJ}. Since the volume of particles is conserved, the front propagation speed $v_f$ along the direction of impact then is related to the impactor speed  $v_p$ via  \cite{dozer}
\begin{equation}
v_f = \frac{\phi_J}{\phi_J-\phi_0} v_p, 
\label{eq:plough}
\end{equation}
where $\phi_J$ is the packing fraction at which jamming occurs and $\phi_0 < \phi_J$ is the packing fraction of the initially unjammed suspension at rest. The closer the initial packing fraction is to jamming, the faster the front will propagate, in principle diverging at $\phi_J$. This model shows excellent agreement with measurements of $v_f$ in systems where the local packing fraction can change easily, such as dry granular particle layers that are being compacted snowplough-like from one end \cite{dozer}.

In suspensions the presence of an interstitial liquid makes it possible to prepare three-dimensional systems at packing fractions $\phi$ well below $\phi_J$ by density matching the particles to the liquid. Given that such systems can still exhibit impact-induced solidification, jamming by densification would imply significant  particle packing fraction changes $\Delta \phi = \phi_J - \phi$.  However, unless the impact speed is so high  that the liquid becomes compressible \cite{Petel}, viscous drag will counteract any densification of the particle sub-phase. This calls into question the mechanism underlying Eq.~1, even though there is experimental evidence for the basic outcome, namely that the ratio $v_f/v_p$ increases dramatically as $\Delta \phi$ approaches zero \cite{Scott,Ivo}. 

One intriguing alternative mechanism has recently emerged with the concept of jamming by shear \cite{SJ}. In this extension of frictionless, standard jamming, the presence of frictional interactions between particles makes it possible to start from initially isotropic, unjammed configurations at $\phi = \phi_0 < \phi_J$ and, without changing $\phi$, rearrange the particles into anisotropic jammed configurations by applying shear. Shear jamming is also possible in frictionless systems, albeit over a much smaller range in $\Delta \phi$ \cite{Kumar}. So far, such shear jamming has been observed experimentally in two-dimensional (2D) dry granular systems under quasi-static conditions, where there is direct visual access to particle positions and stresses by imaging perpendicular to the 2D plane \cite{SJ}. Investigating the role of shear jamming in dynamic impact-induced solidification of 3D suspensions requires the capability of non-invasively tracking the jamming fronts and the associated, quickly evolving flow field in the interior of an optically opaque system.

Here we achieve this with ultrasound. Measuring the speed of sound $c$ we obtain an upper bound on the change of packing fraction $\Delta \phi$ as the suspension jams. We find that at $v_p \ll c$, $\Delta \phi$ is much smaller than required if densification was the primary driver for impact-activated solidification. To investigate the crossover as $v_p$ increases, we use high-speed ultrasound imaging to track the emergence of concentrated shear bands at the location of the propagating jamming fronts. In the regime of small $v_p$, the suspension responds to stress like a fluid, and in the regime of large $v_p$, the suspension develops solid-like characteristics, which we identify by investigating the flow fields. The invariant packing fraction and existance of shear bands provides direct evidence of dynamic shear jamming in 3D suspensions. Furthermore, access to the full flow field allows us to extract the local shear rates and identify the origin of a key, but so far unexplained, feature of the response to impact: the longitudinal front propagation speed exceeds the transverse propagation speed by a factor very close to two \cite{Ivo}.
\\

\noindent
\textbf{Results}


\begin{figure*}
\begin{center}
\includegraphics[scale=0.6]{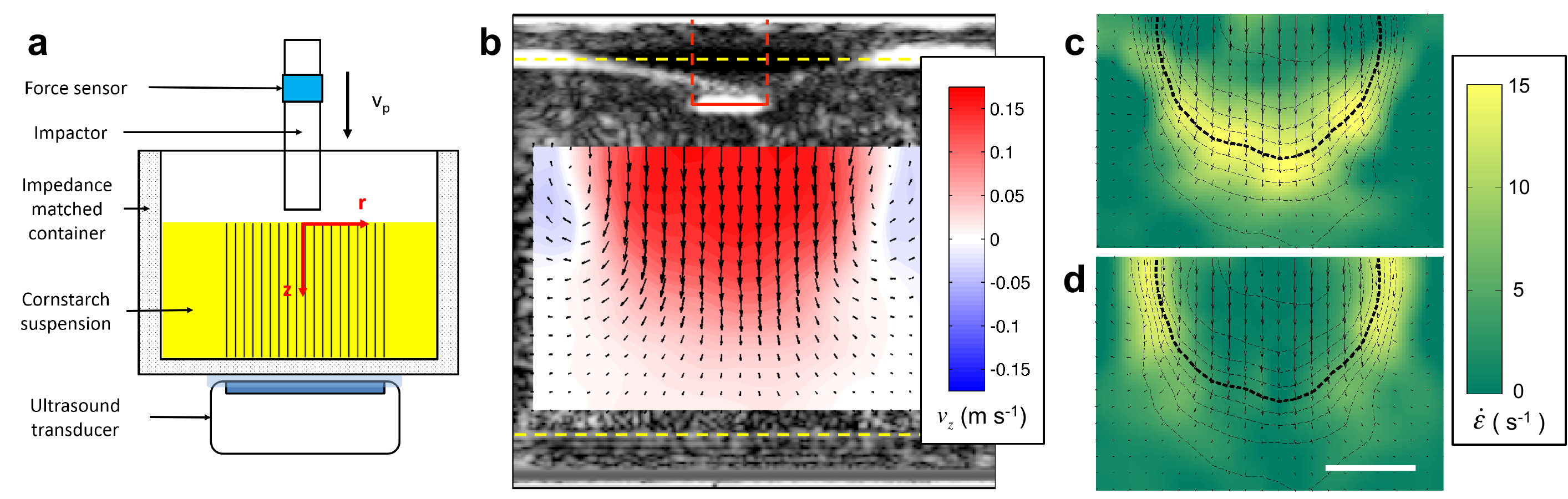}
\end{center}
\caption{\label{Flow} \textbf{Visualization of the flow field with ultrasound.} (a) Sketch of the experimental setup. The sample container and impactor are cylindrical and concentric. The ultrasound transducer scans a vertical slice centered along the central $z$-axis, providing a field of view as indicated by the shaded area. (b) Velocity field during impact. The image is a snapshot from a high-speed ultrasound movie (shown in gray scale) with overlaid velocity field from particle image velocimetry (PIV) analysis. The color code represents the magnitude and direction of the vertical component of the local velocity $u_z$ (red corresponds to downward, blue to upward flow). Dashed yellow lines indicate the locations of the free surface of the suspension and of the bottom of the container. The impactor is outlined in red. The experimental parameters are $\phi = 0.47$, $v_p = 175$~mm$\cdot$s$^{-1}$, liquid viscosity $\eta_0 = 4.6 \pm 0.2$ mPa$\cdot$s, fill depth $H = 30$~mm, and impactor diameter $D = 6.0$~mm. (c), (d) Two components of the shear rate tensor, $|\dot{\varepsilon}_{zz}|$ (c) and  $|\dot{\varepsilon}_{rz}|$  (d), shown for the same instant in time as the flow field in (b). Dashed lines are contours connecting points with the same $u_z$. The thicker line indicates $u_z = v_p/2$, which defines the front position. Scale bar is $1$~cm for (c) and (d). The whole process is shown in Supplementary Movie 1. }
\end{figure*}

\noindent
\textbf{Experimental setup and extraction of the flow field.}
The experiments were performed with a prototypical suspension: cornstarch particles dispersed in water-glycerol CsCl solutions.  The experimental setup is illustrated in Fig.~\ref{Flow}a. In the impact experiments the impactor was driven  vertically downward with constant velocity $v_p$ by a linear actuator. 
A representative flow field $(u_r, u_z)$ inside the suspension during impact is shown in Fig.~\ref{Flow}b. The vertical and horizontal axes in the image correspond to the $z$ and $r$ directions in  cylindrical coordinates. The flow field shows a jammed solid plug in the center, as evidenced by the fact that all points move vertically with speed close to $v_p$. 
Also evident  is a strong velocity gradient around the periphery of the jammed region. To show this more explicitly, we  calculate the local shear rate from the velocity field $(u_r, u_z)$. Given rotational symmetry, the shear rate tensor becomes
\begin{equation}
\dot{{\bf \varepsilon}} = 
\begin{bmatrix}
  \frac{\partial{u_r}}{\partial{r}} & 0 & \frac{1}{2}(\frac{\partial{u_r}}{\partial{z}}+\frac{\partial{u_z}}{\partial{r}}) \\
 0 & \frac{u_r}{r}  & 0 \\
  \frac{1}{2}(\frac{\partial{u_r}}{\partial{z}}+\frac{\partial{u_z}}{\partial{r}}) & 0 & \frac{\partial{u_z}}{\partial{z}} 
\end{bmatrix}
. 
\label{eq:SR_tensor}
\end{equation}
Figures~\ref{Flow}c-d show the two components $|\dot{\varepsilon}_{zz}| = -\frac{\partial{u_z}}{\partial{z}}$ and $|\dot{\varepsilon}_{rz}| = -\frac{1}{2}(\frac{\partial{u_r}}{\partial{z}}+\frac{\partial{u_z}}{\partial{r}})$. Underneath the jammed region, {\sl i.e.}, in longitudinal direction, $\dot \varepsilon_{zz}$ dominates. This corresponds to pure shear that compresses the suspension in the $z$ direction and expands it radially. By contrast, along the sides of the jammed plug $\dot \varepsilon_{rz}$ dominates, and here the main contribution arises  from the term $\frac{\partial{u_z}}{\partial{r}}$. As a result, the velocity gradient is mainly perpendicular to $v_p$. This is analogous to simple shear as seen, for example, in parallel plate setups. We will return below to the implications of having both types of shear .  \\


\begin{figure}[!t]
\begin{center}
\includegraphics[scale=1]{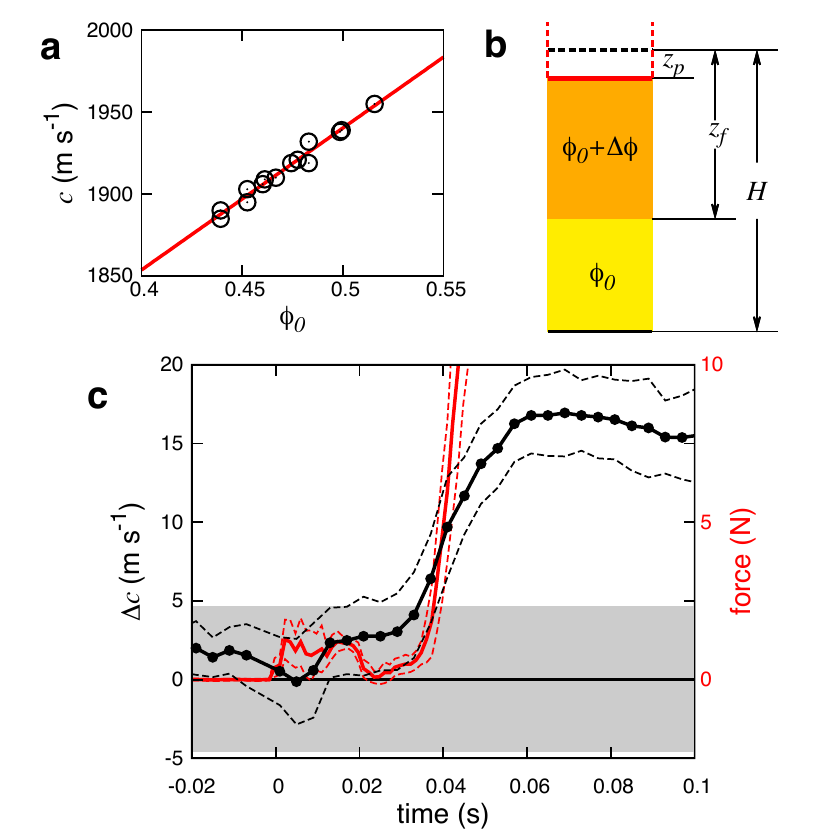}
\end{center}
\caption{\label{SOS} \textbf{Direct measurement of packing fraction changes.} (a) Speed of sound $c$ as a function of packing fraction $\phi$. All data were taken with suspensions in their quiescent fluid-like state, at packing fractions well below $\phi_J$. Here we use data from multiple measurements at the same point to show the overall uncertainty of this experiment. (b) Sketch of the region beneath the impactor. The black dashed line represents the initial suspension surface at a fill height $H$ above the bottom of the container (bold black solid line). As the impactor (outlined in red) pushes down a vertical distance $z_p$ the front (orange region) propagates a distance $z_f$. (c) Change in sound speed $\Delta c$ as a function of time (black trace). Impact at the free suspension surface occurs at $t=0$ ms. Once the jamming front has reached the bottom of the container, all material below the impactor has been transformed into a shear-jammed solid. At that point, near $t= 35$ ms, the force on the impactor (red trace) rises dramatically. Note that within our experimental uncertainties, the speed of sound does not increase until the shear-jammed solid becomes compressed. Data are averages from seven experiments that simultaneously measured force and sound speed as functions of time. Dashed lines indicate one standard deviation. The grey region shows the uncertainty  (given by one standard deviation) in determining $\Delta c$ at low $v_p$, where no solidification takes place. }
\end{figure}

\noindent
\textbf{Speed of sound.}
In an unjammed suspension of solid particles in a Newtonian liquid the shear modulus vanishes. In the limit that the solid particles are much smaller than the  wavelength, the speed of sound is then given by $c = (K/ \rho)^{\frac{1}{2}}$, where $K$ is the average bulk modulus and $\rho$ the average material density of the suspension \cite{Urick, McClements}. In our experiments the particles and suspending solvent are density matched (see Methods), but the average $K$ still increases with $\phi$ since cornstarch particles have a bulk modulus larger  than that of the  liquid \cite{Johnson}. As shown in Fig.~\ref{SOS}a, the resulting dependence of $c$ on $\phi$ is, to a good approximation, linear across the regime of packing fractions probed by our experiments. 
This allows us to obtain $\Delta \phi$ straightforwardly by detecting changes in $c$ with ultrasound. 

A schematic illustration of the suspension under impact is shown in Fig.~\ref{SOS}b, indicating two regions: a jammed region (orange) directly underneath the impactor and an unjammed region (yellow) ahead of the jamming front. Our measurements provide the average speed of sound $\bar c$ as determined from the time it takes sound pulses to traverse both regions (see Methods). The process of transforming unjammed suspension to jammed suspension could be expected to increase the speed of sound via several mechanisms: (1) an increase $\Delta \phi$ in packing fraction; (2) an increase in effective bulk modulus $K$; (3) the development of a finite shear modulus $G$ as the suspension jams \cite{jamming}.

Figure~\ref{SOS}c shows the measured change in sound speed $\Delta c = \bar c - c(\phi_0)$ during impact for a suspension prepared at $\phi_0 = 0.48$. The impactor hits the suspension surface at time $t$ = 0, generating a jamming front that reaches the bottom at $t \approx 0.035$ s.  We can identify this point by the dramatic increase in force on the impactor, as established by prior investigations of quasi-2D \cite{Ivo} and 3D \cite{Scott} systems.  Until the jamming front interacts with the bottom wall $\Delta c = \bar c - c(\phi_0)$ is less than the measurement uncertainty of about $5$~m/s. This allows us to obtain upper bounds on the increase in either packing fraction or moduli associated with the impact-induced solidification process. Neglecting any increases in $K$ and $G$, we find from Fig.~\ref{SOS}a  that our noise floor $\Delta c \approx$ 5m/s implies $\Delta \phi \approx 0.006$ during the free propagation of the fronts. This means that $\phi$ could have increased to 0.49 at best. As shown in Fig.~\ref{SOS}(a), we measured the speed of sound in the suspension at $\phi = 0.52$. At this packing fraction the suspension can still flow when sheared slowly, so $\phi_J$ should be greater than 0.52, which means that the increase in packing fraction due to impact is much less than the prediction of the densification model. If instead we assume $\Delta \phi$ = 0 and use $c = [(K+\frac{4}{3}G)/\rho]^{\frac{1}{2}}$, as appropriate for solids \cite{ultrasound}, to describe the dependence of the speed of sound on $K$ and $G$ within the jammed region behind the front, the same noise floor $\Delta c \approx 5$~m/s implies that the net increase in the sum of the moduli $\tilde K = K+\frac{4}{3}G$ could not have been larger than $\Delta \tilde K \approx 32$~MPa. It is very small compared to the bulk modulus $K_0$ of the quiescent suspension at $\phi_0 = 0.48$: $\Delta \tilde K/K_0 \approx 0.5 \%$. 

Once the front has reached the bottom, $\Delta c$ increases to $\approx 16$~m$\cdot$s$^{-1}$. While this is significantly above the noise floor, it limits any packing fraction changes to  $\Delta \phi \approx 0.02$, still less than necessary to reach $\phi_J$.  Along the same lines, any net increase in moduli would be limited to $\Delta \tilde K \approx 101$~MPa and $\Delta \tilde K/K_0 \approx 1.7 \%$. 
\\

\begin{figure}[!t]
\begin{center}
\includegraphics[scale=1]{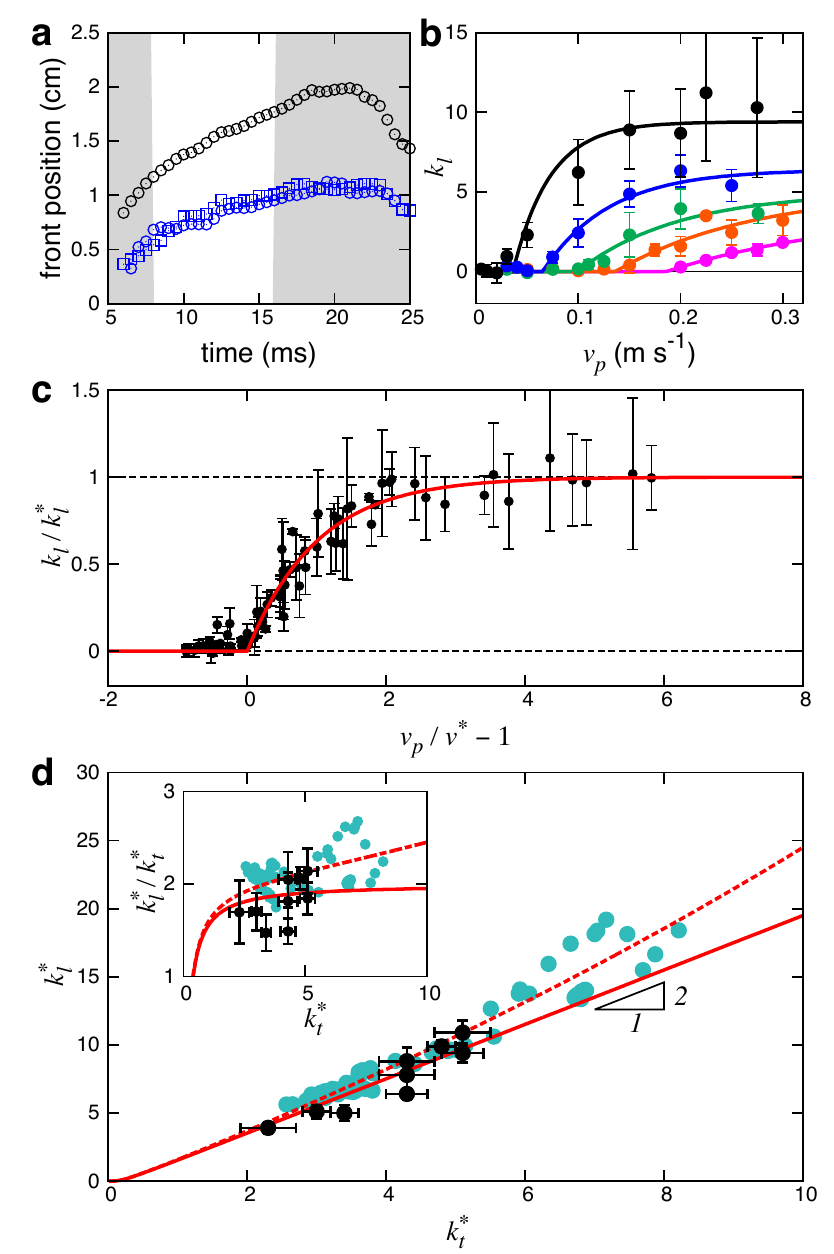}
\end{center}
\caption{\label{kFactor} \textbf{Propagation of  jamming fronts.} 
(a) Front position in  longitudinal (black) and transverse (blue circles: right, blue squares: left) directions as functions of time. The impactor touches the suspension surface at time $t= 0$ ms. The grey shaded background indicates the initial front build-up (left), and the regime where the fronts starts to interact with boundaries and slows down (right). For these data $v_p$ = 200~mms, $\phi$ = 0.460. 
(b) Normalized front propagation speed $k_l$ in longitudinal direction as function of $v_p$ for different $\phi$ (magenta: 0.439, orange: 0.453, green: 0.460, blue: 0.474, black: 0.498). All data are for suspending liquid viscosity $\eta_0 = 4.6 \pm 0.2$~mPa$\cdot$s. Error bars show the standard deviation of three measurements. The same data plotted in log-linear scale is shown in Supplementary Fig. 3. 
(c) Front speed $k_l/k^*_l$ normalized by its asymptotic value as function of impactor speed $v_p$ normalized by threshold speed $v^*$. Data from experiments with different $\phi$ and $\eta_0$ collapse onto a master curve fit by Eq.~\ref{eq:k_fit} (red solid line). 
(d) Relationship between the asymptotic front speeds  $k^*_l$ and $k^*_t$ in longitudinal and transverse direction, respectively. Data from both quasi-2D \cite{Ivo} (turquoise) and 3D (black) impact experiments are shown. The solid red line is the prediction from Eq.~\ref{eq:k_theory}. The dashed red line is a modified version of the model, which includes a small strain anisotropy $\delta$, here plotted using a value of $\delta = 0.01$. Error bars are the asymptotic standard error from the fittings of each $k$-$v_p$ curve with Eq.~\ref{eq:k_fit}. }
\end{figure}

\noindent
\textbf{Propagation of fronts.}
From the evolution of the flow fields as shown in Fig.1 we extract the position of the moving jamming front, defining the front position as the line of points where the vertical component of the impact velocity has dropped to $v_p/2$. In the following we focus on the points in the flow field that propagate the furthest in $z$ and $r$ directions, {\sl i.e.}, on the maximum longitudinal and transverse front speeds. As shown in Fig.~\ref{kFactor}a, after an initial stage the fronts in both directions propagate essentially linearly as function of time before slowing down when they start to interact with boundaries and the incipient jammed solid gets compressed by the impactor; further compression causes the motion to stop quickly (see Supplementary Fig. 1). Here we investigate this linear regime, where the front propagates freely.  
To compare how quickly the front propagates relative to $v_p$, we define two dimensionless front propagation factors, or normalized front speeds, $k$ as 
\begin{equation}
k_t = \frac{v_{ft}}{v_p}, ~ k_l = \frac{v_{fl}}{v_p}-1, 
\label{eq:k_def}
\end{equation} 
where the subscripts $t$ and $l$ represent transverse and longitudinal directions, respectively. The ``$-1$'' in $k_l$ compensates for the vertical motion of the impactor itself. 

Our experiments show that the parameters that affect $k_l$ and $k_t$ include $\phi$, $v_p$, and the suspending liquid's viscosity $\eta_0$. For a suspension with given $\eta_0$ and $\phi_0$ that is impacted very slowly, the response is fluid-like and both $k_t$ and $k_l$ are close to zero.  However, beyond a threshold value $v^*$ jamming fronts start to appear. Their normalized speed initially increases quickly with impactor speed $v_p$ but eventually asymptotes to a fixed $k^*$. The relation between $k_l$ and $v_p$ in suspensions with the same $\eta_0$ but different $\phi_0$ is shown in Fig.~\ref{kFactor}b; the behavior of $k_t$ is similar. As $\phi$ increases the curves shift towards lower $v^*$ and higher $k^*$. For comparison, in suspensions with the same $\phi_0$, larger solvent viscosity $\eta_0$ leads to lower threshold $v^*$ (see Supplementary Fig. 2). 
Plotting the data in terms of normalized variables $k/k^*$ and $v_p/v^*$ scales out the dependencies on $\phi_0$ and $\eta_0$.  In order to extract $k^*$ and $v^*$
we fit the data to the phenomenological relation
\begin{equation}
\frac{k}{k^*} = 
\left \{
\begin{aligned}
& 0 ~ &(v_p \le v^*), \\
& 1-e^{1-v_p/v^*} ~ &(v_p > v^*). \\
\end{aligned}
\right.
\label{eq:k_fit}
\end{equation} 
The resulting data collapse for the longitudinal front speed ratio $k_l/k^*_l$ is plotted in Fig.~\ref{kFactor}c. A similar result is obtained for the transverse speed ratio $k_t/k^*_t$. 
To quantify the anisotropy in front propagation in longitudinal and transverse directions, we plot $k^*_l$ versus $k^*_t$ in Fig.~\ref{kFactor}d, using data from experiments varying $\phi$ (larger packing fraction increases both $k^*_l$ and $k^*_t$) and $\eta_0$. To a good approximation, all data follow $k^*_l = 2 k^*_t$. Comparison with data obtained for quasi-2D suspension \cite{Ivo} shows excellent agreement as well, except that for higher $k_t$ the ratio $k_l/k_t$ slightly exceeds $2$. \\

\noindent
\textbf{Discussion}

Our data in Fig.~\ref{SOS}c demonstrate that impact-activated jamming of dense suspensions proceeds without significant increase of $\phi$, and certainly without increasing $\phi$ to values close to $\phi_J$.  This rules out earlier explanations based on entering the jammed state via densification of the particle sub-phase \cite{Scott, dozer}. On the other hand, analysis of the flow field shows that the jamming fronts initiated by the impact coincide with the location of the maxima in local shear rate (Fig.~\ref{Flow} c and d).  Together, these two findings provide strong evidence for dynamic shear jamming: the impact triggers propagating jamming fronts that locally create sufficient shear to reorganize particles into (anisotropic) jammed configurations without changing the average packing fraction. There are several implications of the shear jamming scenario for suspensions and several differences from dry granular systems, both of which we discuss next.

To start, we examine the stress. In a dry granular system stress is sustained only via direct contact between particles. By contrast, in a dense suspension stress can also be transmitted without contact via lubrication forces. Thus while in dry granular systems there is only a single characteristic stress scale for entry into the shear-jammed regime \cite{SJ}, for a suspension the situation can be more complex. A number of theoretical models \cite{Cates_2}, simulations \cite{Seto, Einav} and experiments \cite{MicroMechST, Itai, Poon} have recently suggested that lubrication breaks down and particles start to experience frictional interactions beyond a local stress threshold $\sigma_1$. Thus, for stress levels below $\sigma_1$ the suspension behaves liquid-like, while above  $\sigma_1$ the system can be thought to behave more like a frictional granular system, {\sl i.e.}, enter a fragile regime before crossing over into the shear-jammed regime at a second characteristic stress level $\sigma_2$ \cite{SJ}.
Within this picture, we associate the transition at $v^*$ with the situation where the stress levels at the leading edge of the jamming front have reached $\sigma_1$ and are large enough for frictional interactions to occur. Thus, when $v_p < v^*$ the suspension is in the lubrication regime, but when $v_p > v^*$ it transitions into a fragile state with behavior intermediate between solid and fluid \cite{SJ, Cates_1, MarginalMatters}, as frictional contacts start percolating through the system to form a load-bearing network, eventually reaching a solid-like shear-jammed state as $v_p$ increases further (Fig.~\ref{kFactor}c). 

The stress-based argument also provides an explanation of the relaxation or ``melting'' of the jammed region when the impact stops. During front propagation the stress inside the jammed region is sustained by the inertia of the suspension in the shear zone ahead of the jammed region. When the motion of the impactor stops, the shear zone disappears and the stress applied on the boundary of the jammed solid falls below $\sigma_1$, insufficient to sustain frictional interactions between particles and therefore any network of force chains that could generate a yield stress and support a load. As a result, the suspension returns to the lubrication regime.

However, while necessary, the existence of threshold stress levels is not sufficient to explain the asymptotic front speed $k^*$ at high $v_p$ and the seemingly universal anisotropy in front propagation, expressed by the ratio $k^*_l/k^*_t \approx 2$. Particles also need to move out of an initial uniform isotropic distribution and reorganize under shear into anisotropic structures (force chains) that can support the stress. Such reorganization requires a minimum shear strain $\varepsilon_c$ to engage neighboring particle layers. As a result, shear jamming happens only when stress and strain both reach their threshold values. In a quasi-static granular system \cite{SJ, Kumar} the threshold strain only matters when the shear jammed state is prepared or when the shear is reversed. In the dynamic system considered here, the front continues to propagate into unperturbed suspension, and therefore, the front advances by applying strain  $\varepsilon_c$ \emph{locally} during the whole process of front propagation. 

For dense suspensions in the high $v_p$ regime, where the stress threshold is clearly exceeded, we can show that $k^*$ is governed by $\varepsilon_c$ and that, in fact, the front speed anisotropy is a direct consequence of the existence of a strain threshold. As described above, the suspension experiences pure shear in the longitudinal direction and simple shear in the transverse direction. In 2D we can directly compare the two types of shear using the positive eigenvalues of the shear rate tensors, treating the propagation of the front in the longitudinal and transverse directions as two effectively 1D problems. We now assume that a suspension element jams when the shear strain it experiences reaches $\varepsilon_c$, irrespective of propagation direction. This leads to the following relations between $k^*_l$, $k^*_t$ and $\varepsilon_c$  (see Methods):
\begin{equation}
k^*_t = 1/(2\varepsilon_{c}), ~ k^*_l = 1/(e^{\varepsilon_{c}}-1)
\label{eq:k_theory}
\end{equation} 
and
\begin{equation}
k^*_l = \frac{1}{e^{1/(2k^*_t)}-1}. 
\label{eq:k_ratio}
\end{equation} 
Equation~\ref{eq:k_ratio} is plotted in Fig.~\ref{kFactor}d. For small $\varepsilon_c$ we find from Eq.~\ref{eq:k_theory} that $k^*_l \approx 2k^*_t = 1/\varepsilon_{c}$. In other words, the anisotropy ratio of $2$ in the normalized front speeds originates from the factor $1/2$ in the non-diagonal terms of the shear rate tensor, which in turn arises because simple shear can be decomposed into a combination of pure shear and solid body rotation. 
In 3D it is not possible to quantify the effects of pure shear and simple shear via the same approach (see Methods). However, one possible solution is to use the ``strain intensity'' $D$ suggested by Ramsay and Huber \cite{Ramsay}, which provides a scalar measure of the relative strength of the two types of shear. As we show in the Methods section, this leads to a ratio  $k^*_l/k^*_t \approx 3/\sqrt 2 \approx 2.12 $, very close to the value for the 2D case. Therefore, an anisotropy ratio $\approx 2$ agrees very well with the experimental data for both quasi-2D \cite{Ivo} and 3D systems within our measurement precision. 

With increasing packing fraction $\phi$ the average distance between particles decreases and we expect the strain threshold $\varepsilon_c$ to decrease as well. Via Eq.~\ref{eq:k_theory} this explains the increase in $k^*$ with $\phi$ seen in Fig.~\ref{kFactor}b and also agrees qualitatively with observations in dry granular systems \cite{SJ}: since it takes less strain to reorganize the particles into a shear-jammed network the front will propagate faster for given impactor velocity. We point out that Eq.~\ref{eq:plough}, which formalizes such relationship between packing fraction and front speed, appears to capture the overall trend qualitatively. However, this seems fortuitous, since Eq.~\ref{eq:plough} was based on the assumption that the moving front significantly increases the packing fraction, in fact driving it up all the way to $\phi_J$, which we now can rule out. In addition, Eq.~\ref{eq:plough} cannot predict the observed propagation anisotropy. One of the outstanding tasks therefore is to develop a model for $k^*(\phi)$ that is based on jamming by shear rather than densification. 

An interesting aspect of the data in Fig.~\ref{kFactor}d is the deviation from the anisotropy ratio $\approx$ 2 at large $k^*$ values or, equivalently, large packing densities.  This is most apparent in the data available for the quasi-2D system and it indicates that the longitudinal speed becomes faster. We speculate that this may be connected to a breakdown of the assumption of an isotropic threshold $\varepsilon_c$. For example, if the impact were to introduce a small amount of compression of the particle sub-phase in longitudinal direction, $\varepsilon_c$ would be reduced in that direction. This effect would become increasingly significant at large $\phi$. 
We can model this by introducing a correction $\delta$ so that 
\begin{equation}
k^*_l = 1/(e^{\varepsilon_{c}-\delta}-1). 
\label{eq:k_modified}
\end{equation}
Using $\varepsilon_c = 1/(2 k^*_t)$ and $\delta \approx 0.01$ we can reproduce the trend in Fig.~\ref{kFactor}d well (dotted red line). However, we point out that this is just the simplest way to account for the trend in the data and there might be other reasons for the deviation. 

Taken together, these results provide important insights into the mechanism responsible for impact-induced solidification of dense suspensions. The finding that the packing fraction does not increase measurably during impact, together with the observation of strong shear at the leading edge of the propagating solidification fronts, rules out jamming via densification as the dominant mechanism and points to jamming by shear (densification is likely to play a significant role at much larger impact velocities, when the interstitial liquid's compressibility no longer can be neglected \cite{Petel}). In dense suspensions this introduces a new stress scale or, equivalently, an impact velocity threshold, which we associate with the breakdown of lubrication films between particles and the onset of frictional interactions. Behind the front, these frictional interactions create a dynamically shear-jammed region. This rapidly growing shear-jammed region behaves like a solid, for example when it interacts with a system boundary. Further support for the shear-jamming scenario comes from the observation of anisotropic front propagation, where we can relate the fact that the fronts propagate longitudinally twice as fast as in transverse direction to an equivalent factor in the ability to transmit shear strain. We point out that for dynamic shear jamming, both shear stress and strain need to exceed threshold values, and the critical shear strain determines the front propagation speed. 
\\

\noindent
\textbf{Methods}
\small

\noindent
\textbf{Experimental setup.}
The ultrasound measurements and imaging were performed with a Verasonics Vantage 128 system. The sample container was 3D-printed from UV-cured resin (``Vero White Plus'', Objet Geometries Inc.) whose acoustic impedance matched the suspensions we studied. The inner diameter of the container was $10.0$~cm and the typical depth $H$ of the suspension was $2.5$ to $3.5$~cm. This insured that the front reached the bottom before it interacted with the side wall. The impactor was a cylinder of diameter $D$ of $6$~mm or $10$~mm, driven by a computer controlled linear actuator (SCN5, Dyadic Systems) and equipped with a force sensor (DLC101, Omega). The ultrasound transducer (Philips L7-4; $128$ independent elements; total width of $3.8$~cm) contacted the bottom of the container through a thin layer of ultrasound gel. The ultrasound system was triggered as the impactor approached the surface; images were taken at a frame rate of 10 to 10,000 frames per second, adjusted according to $v_p$. The spatial resolution of the ultrasound was limited by the wavelength, about 0.4mm in our experiments. 
\\

\noindent
\textbf{Suspensions.}
The cornstarch (Ingredion) was stored in the lab environment at $22.5 \pm 0.5~^\circ$C and $51 \pm 2 \% $ relative humidity. Individual particles had a diameter of 5 - 30 $\mu$m \cite{Scott, Starch}. The suspending solvent was a solution of CsCl, glycerol and water. Its viscosity $\eta_0$ was adjusted by the mass ratio of glycerol and water, and its density was matched to that of the cornstarch particles by adjusting the mass ratio of CsCl. The density of the particles was $\rho_{cs} = (1.63 \pm 0.01) \times 10^3$~kg~m$^{-3}$ as measured by density matching. To calculate the packing fraction $\phi$, an accurate determination of the volume occupied by particles and interstitial liquid is required. For cornstarch this is difficult because the particles are porous and they already contain some moisture before they are dispersed in the suspending solvent. Therefore, often a value $\phi_m$ based on the mass fractions before mixing is quoted \cite{DST_review, Scott, Shomeek}, which is proportional to $\phi$. For example, in Figs.~\ref{kFactor}b, the $\phi_m$ values were 0.390, 0.402, 0.409, 0.425, and 0.444. To obtain the actual packing fraction, we account for a small percentage $\alpha$ of suspending liquid that is wicked up by the porous particles and write $\phi = (1+\alpha) \phi_v$, where $\phi_v$ is the material volume packing fraction \cite{CS_1, CS_2}. When calculating $\phi_v$, we considered the moisture in the cornstarch particles. We assume the moisture is pure water and its mass ratio in cornstarch is $\beta$ in our lab environment. This leads to 
\begin{equation}
\phi_v = \frac{(1-\beta)m_{cs}/\rho_{cs}}{(1-\beta)m_{cs}/\rho_{cs} + m_l/\rho_l + \beta m_{cs}/\rho_w}, 
\label{eq:phi}
\end{equation}
where $m_{cs}$ is the mass of cornstarch particles, $m_l$ is the mass of the solvent liquid, $\rho_l$ is the density of the solvent and $\rho_w$ is the density of pure water. 
In this paper we use $\alpha = 0.3$ and $\beta = 0.11$ by estimation. Changing of these numbers will not affect the conclusions we make.   

Since air bubbles mixed in the suspension scatter ultrasound signals significantly, the suspensions were debubbled before using. To keep $\phi$ fixed during debubbling, we filled the suspensions into sealed syringes and then lowered the pressure by pulling the plungers. The syringe walls were tapped gently to help bubbles separate from the suspension. After debubbling a small amount of suspension was used to measure the speed of sound $c$ as required for image reconstruction. For imaging the flow field, a small amount of air bubbles were added back to the debubbled suspensions to act as tracer particles. This was done by slowly stirring the suspension, then tilting and slowly rotating the container till the bubbles were uniformly distributed. The amount of bubbles was small enough so that they did not suppress the penetration of the ultrasound in the suspensions and the change in speed of sound was negligible for our measurements. We also determined that the effect of the bubbles on the suspension viscosity is limited (see Supplementary Fig. 4). 
Between successive impact experiments the suspension was relaxed by gently shaking and rotating the container. 
\\

\noindent
\textbf{Data acquisition and analysis.}
Once triggered, the ultrasound system made several hundred acquisitions consecutively. In one acquisition each of the 128 transducer elements transmitted the same ultrasonic pulse at the same time and received an individual reflected time series. The pulse was a 5 MHz sinusoidal wave modulated by a Gaussian profile for a pulse length of 6 periods. From the time series received by the transducer and using the previously measured speed of sound $c$ we reconstructed B-mode images (using brightness to represent the echo signal amplitude) \cite{ultrasound} that captured the positions of the tracer particles (air bubbles) in the suspension. Given our finding that $c$ does not change measurably during the impact, the image reconstruction process does not need to account for spatial or temporal variations in $c$. By tracking the tracer bubble displacements with a particle imaging velocimetry (PIV) algorithm, we obtained a two-dimensional flow field from within the bulk of the suspension. 
\\

\noindent
\textbf{Change of packing fraction measurements.}
The experimental setup was identical to the one illustrated in Fig.~\ref{Flow}a and a schematic illustration is shown in Fig.~\ref{SOS}b. The impactor started from the surface of the suspension and pushed down a distance $z_p$. The position of the impactor was measured with a high speed camera (Phantom V9, Vision Research). The ultrasound measured the time of flight $T$ of the signal transmitted from the bottom, reflected by the impactor and sent back to the bottom. 
Thus the average speed of sound $\bar c$ along this path is
\begin{equation}
    \bar c = 2 \frac{H-z_p}{T}. 
\label{eq:c_mean}
\end{equation}

We started with experiments at a low $v_p$ ($5$~mm$\cdot$s$^{-1}$) to measure the speed of sound in the liquid-like, unjammed suspension, where $\bar c = c(\phi_0)$. Define $T_0$ as the initial time of flight when $z_p = 0$~mm, we have $H = c(\phi_0) T_0/2$, and from this 
\begin{equation}
    c(\phi_0) = \frac{2z_p}{T_0-T}.   
    \label{eq:c_0}
\end{equation}
The initial packing fraction $\phi_0$ in these experiments was $0.48$. The liquid was a mixture of $44.3\%$ CsCl, $27.8\%$ glycerol and $27.8\%$ water by mass, with $\eta_0 = 4.6$~mPa$\cdot$s and $\rho = 1.63 \times 10^3$~kg$\cdot$m$^{-3}$. From six measurements we obtain $c(\phi_0) = 1939.2 \pm 4.6$~m$\cdot$s$^{-1}$ and $H = c(\phi_0)T_0/2 = 34.1 \pm 0.1$~mm. 

For the high $v_p$ ($200$~mm$\cdot$s$^{-1}$) experiments we used the value of $H$ measured above and equation~\ref{eq:c_mean} to calculate $\bar c$. The time of flight now becomes
\begin{equation}
    T = 2 \big[ \frac{H-z_f}{c(\phi_0)} + \frac{z_f-z_p}{c(\phi_0+\Delta \phi)} \big].   
    \label{eq:travel_time}
\end{equation}
For $\Delta \phi = 0$, $T_{\Delta \phi = 0} = 2(H-z_p)/c(\phi_0)$. If $\Delta \phi > 0$, there will be a difference between $T$ and $T_{\Delta \phi = 0}$, and the difference becomes increasingly large as $z_f$ increases, which leads to an increase in $\bar c$ according to equation~\ref{eq:c_mean}. 
\\

\noindent
\textbf{Derivation of Equation~\ref{eq:k_theory} in 2D.}
For an idealized 2D system we define a Cartesian coordinate with x axis in the transverse  and y axis in the longitudinal direction. To obtain the relation between the strain threshold $\varepsilon_c$ and the normalized front speeds $k$ we consider how much shear strain a suspension element experiences when it accelerates from $u_y = 0$ to $u_y = v_p$. We consider the propagation in the transverse and longitudinal directions separately as two quasi-1D problems. Exemplary sketches of the velocity profiles are provided in Supplementary Fig.~5.  The experimental data did not show a significant change in front width, so here we assume the shape of the front does not change during propagation. In this case the velocity profiles can be expressed as
\begin{equation}
    u_y(x,t) = f_t(x-v_{ft} t) 
    \label{eq:f_shape}
\end{equation}
in the transverse direction and 
\begin{equation}
    u_y(y,t) = f_l(y-v_{fl} t) 
    \label{eq:f_shape}
\end{equation}
in the longitudinal direction.
In both equations $t$ is time, $v_{ft}$ and $v_{fl}$ are front propagation speeds. $f_t(X)$ and $f_l(X)$ are functions that satisfy $f_t = f_l = v_p$ as $X \to -\infty$ and  $f_t = f_l = 0$ as $X \to + \infty$. 

On either side of the impactor the front propagates in transverse direction and the front speed $v_{ft} = k_t v_p$, while the suspension itself is sheared longitudinally by the advancing front. The acceleration of a suspension element is then
\begin{equation}
    \frac{Du_y(x,t)}{Dt} = \frac{\partial f_t}{\partial t} = -k_t v_p f_t' = -k_t v_p \frac{\partial u_y}{\partial x},  
    \label{eq:a_t}
\end{equation}
where $D/Dt$ is the material derivative and $f_t' = df_t(X)/dX$. Below the impactor there are two differences: one is that the suspension element now moves along the propagation direction of the front and the other is that  $v_{fl} = (k_l+1)v_p$ as defined in Eq. 3. The acceleration then becomes
\begin{equation}
\begin{aligned}
    \frac{Du_y(y,t)}{Dt} &= \frac{\partial f_l}{\partial t} + (u_y\frac{\partial}{\partial y})f_l = [u_y - (k_l+1)v_p] f_l' \\
	&= [u_y - (k_l+1)v_p] \frac{\partial u_y}{\partial y}. 
   \label{eq:a_l} 
\end{aligned}
\end{equation}

Now we look at the relation between the local shear rate $\dot{\varepsilon}$ and the velocity gradient.  In general, for an incompressible 2D fluid the shear rate tensor is
	$$
	\dot{\boldsymbol \varepsilon} = 
	\begin{bmatrix}
 	\frac{\partial u_x}{\partial x} &  \frac{1}{2} (\frac{\partial u_x}{\partial y} + \frac{\partial u_y}{\partial x}) \\
 	\frac{1}{2} (\frac{\partial u_x}{\partial y} + \frac{\partial u_y}{\partial x}) & \frac{\partial u_y}{\partial y} 
	\end{bmatrix},
	$$
where $ \frac{\partial u_x}{\partial x} = -\frac{\partial u_y}{\partial y}$. From experimental observation we have $\frac{\partial u_x}{\partial y} \ll \frac{\partial u_y}{\partial x}$. For the transverse direction, where simple shear dominates, the diagonal terms vanish and the shear rate tensor becomes 
	$$
	\dot{\boldsymbol \varepsilon}_t = 
	\begin{bmatrix}
 	0 & \frac{1}{2} \frac{\partial u_y}{\partial x} \\
 	\frac{1}{2} \frac{\partial u_y}{\partial x} & 0 
	\end{bmatrix},
	$$
while for pure shear in the longitudinal direction the off-diagonal terms vanish and we have
	$$
	\dot{\boldsymbol \varepsilon}_l = 
	\begin{bmatrix}
 	-\frac{\partial u_y}{\partial y} & 0 \\
 	0 & \frac{\partial u_y}{\partial y} 
	\end{bmatrix}.
	$$
In either case the matrix has two eigenvalues with the same magnitude but opposite sign and the eigenvalues represent the shear rate on the principal axes. Thus we can represent the shear intensities by the tensors' positive eigenvalues: $\dot{{\bf \varepsilon}}_l = |\frac{\partial u_y}{\partial y}| = -\frac{\partial u_y}{\partial y}$ and $\dot{{\bf \varepsilon}}_t = \frac{1}{2} |\frac{\partial u_y}{\partial x}| = -\frac{1}{2} \frac{\partial u_y}{\partial x}$. 

Using the velocity gradient, we relate the local shear rate with the acceleration of the element:
\begin{equation}
    \dot{\varepsilon}_t = \frac{1}{2} \frac{1}{k_t v_p} \frac{Du_y}{Dt} ,  
    \label{eq:dstrain_t}
\end{equation}
\begin{equation}
    \dot{\varepsilon}_l = \frac{1}{(k_l+1)v_p-u_y} \frac{Du_y}{Dt} .  
    \label{eq:dstrain_l}
\end{equation}
Consequently, the total shear strain $\varepsilon$ an suspension element experiences before jamming is: 
\begin{equation}
    \varepsilon_t = \int_0^\infty \dot{\varepsilon}_t dt = \int_0^{v_p} \frac{1}{2k_t v_p} du_y =  \frac{1}{2k_t},   
    \label{eq:strain_t}
\end{equation} 
and 
\begin{equation}
    \varepsilon_l = \int_0^\infty \dot{\varepsilon}_l dt = \int_0^{v_p} \frac{1}{(k_l+1)v_p-u_y} du_y =  \ln (\frac{k_l+1}{k_l}) .   
    \label{eq:strain_l}
\end{equation} 
Eq.~\ref{eq:strain_l} gives $\varepsilon_l  \approx 1/k_l$ for $k_l \gg 1$. If we assume the strain threshold to jamming $\varepsilon_c$ is isotropic, then $k_t = 1/(2\varepsilon_c)$ and $k_l = 1/(e^{\varepsilon_c}-1)$. 
\\

\noindent
\textbf{Relation between $k_l$ and $k_t$ in 3D.}
In 3D the shear rate tensor is shown in equation~\ref{eq:SR_tensor}. In the longitudinal direction pure shear dominates and the shear rate tensor is 
\begin{equation}
\dot{\boldsymbol \varepsilon}_l = 
\begin{bmatrix}
  \frac{\partial{u_r}}{\partial{r}} & 0 & 0 \\
 0 & \frac{u_r}{r}  & 0 \\
 0 & 0 & \frac{\partial{u_z}}{\partial{z}} 
\end{bmatrix}, 
\label{eq:3D_SR_l}
\end{equation}
where $ \frac{\partial{u_r}}{\partial{r}} \approx \frac{u_r}{r} $ and $ \frac{\partial{u_r}}{\partial{r}} + \frac{u_r}{r} +  \frac{\partial{u_z}}{\partial{z}} = 0$. In the transverse direction simple shear dominates. This gives 
\begin{equation}
\dot{\boldsymbol \varepsilon}_t \approx
\begin{bmatrix}
 0 & 0 & \frac{1}{2}\frac{\partial{u_z}}{\partial{r}} \\
 0 & 0 & 0 \\
  \frac{1}{2}\frac{\partial{u_z}}{\partial{r}} & 0 & 0
\end{bmatrix},
\label{eq:3D_SR_t}
\end{equation}
where we have used $\frac{\partial v_r}{\partial z} \ll \frac{\partial v_z}{\partial r}$. Note that though the system is three-dimensional simple shear only operates in the rz plane while leaving the azimuthal direction invariant. The eigenvalues become $\dot{e}_l = \{ -\frac{1}{2}\frac{\partial{u_z}}{\partial z}, -\frac{1}{2}\frac{\partial{u_z}}{\partial z}, \frac{\partial{u_z}}{\partial z} \}$ and $\dot{e}_t = \{ -\frac{1}{2}\frac{\partial{u_z}}{\partial r}, 0, \frac{1}{2}\frac{\partial{u_z}}{\partial r} \}$. Unlike the 2D case, we cannot simply use a positive eigenvalue to represent the shear intensity. 
However, we can define infinitesimal strains $e_i~(i = 1,2,3)$ along the three principal axes and rank-order them according to $ e_1 > e_2 > e_3$. Following the definition given in Ref. \cite{Ramsay}, the ``strain intensity'' $D$ is 
\begin{equation}
\begin{aligned}
D &= \sqrt{\big(ln \frac{1+e_1}{1+e_2} \big)^2 + \big(ln \frac{1+e_2}{1+e_3} \big)^2} \\
&\approx \sqrt{(e_1-e_2)^2 + (e_2-e_3)^2}. 
\label{eq:D}
\end{aligned}
\end{equation}
For pure shear in the longitudinal direction $e_1 = e_2 = -e_3/2$ and $\dot e_3 =  \frac{\partial{u_z}}{\partial z}$, so $D_l \approx \frac{3}{2} |e_3|$, which leads to $\dot D_l \approx -\frac{3}{2}\frac{\partial u_z}{\partial z}$. For simple shear in the transverse direction $e_1 = -e_3$, $e_2 = 0$ and  $\dot e_3 = \frac{1}{2}  \frac{\partial{u_z}}{\partial r}$. This leads to $D_t \approx \sqrt{2} |e_3|$, and therefore $\dot D_t \approx -\frac{\sqrt{2}}{2}\frac{\partial u_z}{\partial r}$. Following the procedure for the 2D case we have 
\begin{equation}
    \dot{D}_t = \frac{\sqrt{2}}{2} \frac{1}{k_t v_p} \frac{Du_z}{Dt} ,  
    \label{eq:Ddot_t}
\end{equation}
\begin{equation}
    \dot{D}_l = \frac{3}{2} \frac{1}{(k_l+1)v_p-u_z} \frac{Du_z}{Dt} .  
    \label{eq:Ddot_l}
\end{equation}
Integration then leads to 
\begin{equation}
    D_t =  \frac{\sqrt{2}}{2} \frac{1}{k_t}, ~ D_l = \frac{3}{2} \ln (\frac{k_l+1}{k_l}) .   
    \label{eq:D}
\end{equation} 
Now we again assume that the system shear-jams when $D$ reaches a threshold strain value $D_c$, independent of the type of shear it experiences. From this we find 
\begin{equation}
k^*_l = \frac{1}{e^{\sqrt{2}/(3k^*_t)}-1}. 
\label{eq:k_ratio_3D}
\end{equation} 
and $k^*_l/k^*_t \approx 3/\sqrt 2 \approx 2.12$ for large $k$. 
\\


\footnotesize

\small
\noindent
\textbf{Acknowledgements}\\
\noindent
We thank Patrick La Riviere for help with the ultrasound system. We thank Mengfei He, Nicole James, Victor Lee, Sayantan Majumdar, Scott Waitukaitis, Tom Witten, and Qin Xu for insightful discussions. 
This work was supported by the US Army Research Office through grants W911NF-12-1-0182 and W911NF-13-10330. Additional support was provided by the Chicago MRSEC, which is funded by NSF through grant DMR-1420709. \\

\noindent
\textbf{Author Contributions}\\
\noindent
All authors designed the experiments and wrote the paper. E.H. performed the experiments and analyzed the data. I.R.P. contributed the PIV code.

\end{document}